\documentclass{pasj01}
\draft 
\begin{document}
\Received{}%{yyyy/mm/dd}
\Accepted{}%{yyyy/mm/dd}

\title{New Scenario of plasma evolution in IC 443}

\author{
Arisa \textsc{Hirayama}\altaffilmark{1 ${\ast}$},
Shigeo \textsc{Yamauchi}\altaffilmark{1},
Kumiko K. \textsc{Nobukawa}\altaffilmark{1},
Masayoshi \textsc{Nobukawa}\altaffilmark{2},
and 
Katsuji \textsc{Koyama}\altaffilmark{3}
}

\altaffiltext{1}{Department of Physics, Nara Women's University, Kitauoyanishimachi, Nara 630-8506}
\altaffiltext{2}{Faculty of Education, Nara University of Education, Takabatake-cho, Nara 630-8528}
\altaffiltext{3}{Department of Physics, Graduate School of Science, Kyoto University, \\
Kitashirakawa-oiwake-cho, Sakyo-ku, Kyoto 606-8502}

\email{raa-hirayama@cc.nara-wu.ac.jp}


\maketitle

\begin{abstract}
Most of young and middle-aged supernova remnants (SNRs) exhibit an ionizing plasma (IP), an ionizing 
process following a shock heated SNR gas. On the other hand, significant fractions of SNRs exhibit a recombining
plasma (RP). 
The origin and the mechanisms of the RP, however,  are not yet well understood.  
This paper proposes a new model that the RP is followed after the IP process taken at the first  epoch of the SNR evolution. 
Using the high quality and wide band (0.6$-$10 keV) spectrum of IC 443, we nicely fitted with a model of two 
RP (two-RP model)  plus a power law (PL) with an Fe\emissiontype{I} K$\alpha$ line component.  
The ionization temperature in one RP monotonously increases from Ne--Ca, while that in the other RP shows a drastic increase from Cr$-$Ni. 
Origin and mechanism of the two-RP and PL with an Fe\emissiontype{I} K$\alpha$ line components are possibly due to a different evolution of two plasmas 
and ionization by the low-energy cosmic ray.

\end{abstract}

\section{Introduction}

The common concept  of a shock heated plasma of supernova remnants (SNRs) is 
that electron temperature ($kT_{\rm e}$) and  ionization temperature ($kT_{\rm i}$) are nearly $\sim$0 keV (atoms are nearly neutral) 
at the initial epoch. 
Soon after, $kT_{\rm e}$ increases quickly, and then gradually ionized atoms, 
or $kT_{\rm i}$ follows after $kT_{\rm e}$. Thus the SNR plasma is not in collisional 
ionization equilibrium (CIE: $kT_{\rm e}$=$kT_{\rm i}$), but is an ionizing plasma (IP: $kT_{\rm e}>kT_{\rm i}$). 
Since X-ray spectra of most of the young-middle aged SNRs are well fitted with  the IP model, 
IP is widely accepted to be a standard SNR evolution (hereafter, IP SNR).
However, recently some SNRs have been known to exhibit a recombining plasma (RP) of $kT_{\rm e} < kT_{\rm i}$. 
(hereafter, RP SNR).  
On the contrary to the clear physics of IP SNRs,  physics for the evolution of RP SNRs or how to make $kT_{\rm e} < kT_{\rm i}$ 
plasma has been so far unclear.  

IC 443 (G189.1$+$3.0) is an SNR with the diameter of 45$'$ at the distance of 1.5 kpc (e.g., \cite{Welsh2003}). 
In the X-ray band, the age is estimated to be 3000--30000 yr (e.g., \cite{Petre1988, Olbert2001}), a middle-old aged SNR. 
It is also reported as a core collapsed (CC-SNR) and mixed morphology SNR (MM-SNR: \cite{Rho1998}). 
A hint of RP was discovered in the IC 443 spectrum for the first time by \citet{Kawasaki2002}.  
\citet{Yamaguchi2009} then discovered an enhanced structure of the radiative recombining continuum (RRC) of the He-like silicon (Si\emissiontype{XIII}) 
and sulfur (S \emissiontype{XV}), which were direct evidence for RP. 
The RRC of the He-like iron (Fe \emissiontype{XXV}), a key element of the evolution, was found  from this middle-old aged SNR 
in the limited energy band of 3.7--10 keV \citep{Ohnishi2014}. 
The RRC structures of these key elements  play a key roll to investigate the spectral evolution of the RP SNRs.  
We, therefore, made a deep observation of IC 443 with Suzaku \citep{Mitsuda2007} to establish the RP structures in many key elements.

Conventionally, a model of the RP process is treated to start from a common  $kT_{\rm i}$ in all the elements. 
We propose a new model of the RP, which starts from element-dependent $kT_{\rm i}(z)$.
For the verification of this new model, and to combine the RP to the well-established IP process, 
we utilize the 0.6--10 keV band spectrum of IC 443 obtained in the $\sim$400 ks exposure observations with Suzaku.  
Details of the new RP model, analysis process and results are presented in section 3.  Based on the results, 
implications for the RP origin in IC 443 are discussed in section 4. 
The quoted errors are in the 90\% confidence limits.

\section{Observations}

The Suzaku observations of IC 443 were performed 
with the X-ray Imaging Spectrometer (XIS: \cite{Koyama2007}) placed at the focal planes of the thin foil X-ray Telescopes (XRT: \cite{Serlemitsos2007}). 
The XIS consisted of 4 sensors.
One of the XIS (XIS1) is a back-side illuminated CCD (BI), while
the other three XIS sensors (XIS0, 2, and 3) are a front-side illuminated CCD (FI). 
In order to achieve the highest count rate ratio of IC 443/NXB in the higher energy band, 
we only used the FI following \citet{Ohnishi2014}. 
The count rate ratio of IC 443/NXB is determined by the count rates of IC 443 and NXB. 
The photon count rate ratio of FI and BI of IC 443 is estimated by their quantum efficiencies, 
and is $\sim$1.5--1.8 at the Fe\emissiontype{XXVI} Ly$\alpha$ and RRC energy band of 7--10 keV (see Fig. 4 of \cite{Koyama2007}), 
while the count rate ratio of NXB is $\sim$0.3--0.06 (see Fig. 16 of \cite{Koyama2007}).  
As a matter of fact, the observed IC 443 photon count rates of FI and BI are 3.0$\times 10^{-3}$ counts s$^{-1}$ and 1.5$\times 10^{-3}$ counts s$^{-1}$, 
while the count rates for NXB are 1.56$\times 10^{-2}$ counts s$^{-1}$ and 6.44$\times 10^{-2}$ counts s$^{-1}$, respectively.  
Therefore, the count rate ratio of IC 443/NXB of FI is $\sim$0.2, about 10 times better than that of BI of $\sim$0.02. 
Accordingly, we only use the FI in the analysis of the high energy band of 7--10 keV.  
The worse count rate ratio of IC 443/NXB of BI than FI is mainly attributable to the larger NXB count rate of BI than FI. The data used in this paper are listed in Table 1. 

The XIS was operated in the normal clocking mode. The field of view (FOV) of the XIS is \timeform{17.'8}$\times$\timeform{17.'8}.
The XIS employed the spaced-row charge injection (SCI) technique to recover the spectral resolution \citep{Uchiyama2009}.
The XIS data in the South Atlantic Anomaly, during the earth occultation, and at the low elevation angle from the earth rim of $<5^{\circ}$ 
(night earth) and $<20^{\circ}$ (day earth) are excluded. 
Removing hot and flickering pixels, the data of the Grade 0, 2, 3, 4, and 6 are used. 
The XIS pulse-height data are converted to Pulse Invariant (PI) channels using the {\tt xispi} software in the HEAsoft 6.19
and the calibration database version 2016-06-07. 
We used only XIS0 and XIS3 after 2006 November because XIS2 became out-of-function after the epoch. 
A small fraction of the XIS0 area became unavailable, possibly due to an impact of micro-meteorite 
on 2009 June 23. After the epoch, we ignore the damaged area  of the XIS0.

\begin{table*}[t] %Table 1
 \caption{Observation logs.}
  \centering
   \begin{tabular}{cccc} \hline
     Obs. ID & Obs. date & (R.A., Dec.)$_{\rm J2000.0}$ & Exposure \\ 
                   & start time -- end time &                                                 & (ks) \\\hline
                   \multicolumn{4}{c}{IC 443} \\ \hline
     501006010 & 2007-03-06 10:40:19 -- 2007-03-07 12:22:14 & (6:17:11, 22:46:32) & 42.0  \\
     507015010 & 2012-09-27 05:29:48 -- 2012-09-29 18:40:22 & (6:17:11, 22:45:12) & 101.8  \\
     507015020 & 2013-03-27 04:15:06 -- 2013-03-28 16:00:19 & (6:17:12, 22:44:47) & 59.3  \\
     507015030 & 2013-03-31 11:44:34 -- 2013-04-03 21:12:21 & (6:17:12, 22:44:46) & 131.2  \\
     507015040 & 2013-04-06 05:21:49 -- 2013-04-08 02:00:21 & (6:17:12, 22:44:52) & 75.6  \\ \hline
                   \multicolumn{4}{c}{Background} \\ \hline
     409019010 & 2014-10-05 15:12:56 -- 2014-10-07 19:09:17& (6:27:15, 14:53:24)& 82.1\\
     \hline
   \end{tabular}
\end{table*}

\section{Analysis and Results}

\subsection{X-ray Image} %section 3.1

X-ray photons of IC 443 are taken from the data in the observations listed in table 1 (1--5 rows).  
After subtraction of the non X-ray background (NXB)  generated by {\tt xisnxbgen} \citep{Tawa2008},
the X-ray images in the 2.2--5.2 keV (color map) and 5.5--10 keV (green contour) bands are shown in figure 1, 
where calibration source regions are excluded. 
These energy bands are selected from ejecta dominant plasma (see figures 3c and 3d).  
In comparison with diffuse structure in the  2.2--5.2 keV band, the  5.5--10 keV band image shows clear concentration 
toward the image center (cross mark).  
This contrast would be closely related to the 2-component spectral structure found in section 3.4.

\begin{figure}[t] %figure 1
\begin{center}
\includegraphics[width=8cm, angle=90]{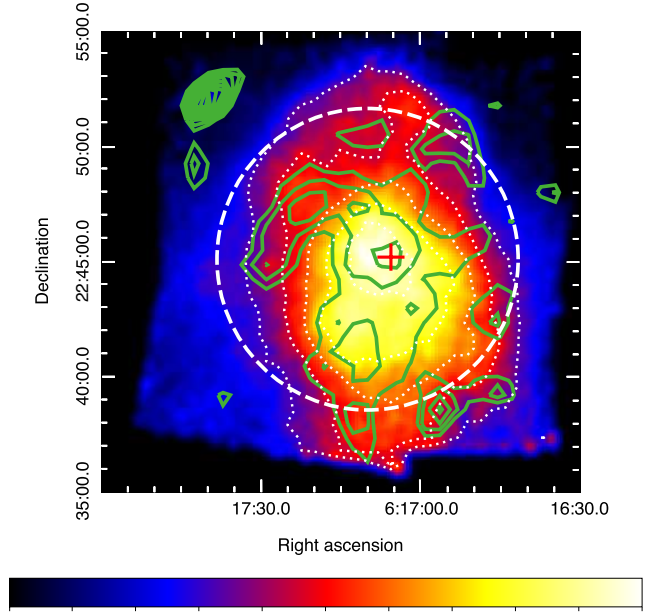} %model A single kTi
\end{center}
\caption{X-ray map of the main part of IC 443 in the 2.2--5.2 keV (color image and white dotted line, RP1) and 5.5--10 keV (contour image, RP2) bands. 
The coordinates are J2000.0.
The subtraction of NXB and the vignetting correction are made. 
The scales are in the linear (peak to bottom). 
The red cross shows a peak of the RP2 component. 
The IC 443 spectrum is taken from the white dashed circle.
The bright source in the 5.5--10 keV band at the upper-left is unrelated point source (No. 9 in table 2 of \cite{Bocchino2003}). }
\end{figure}

\subsection{Overview of the RP Process} %section 3.2

In contrast to the well established IP SNR scenario, that of RP SNR has not been established yet.  In this paper, we propose a viable %natural 
scenario based on the standard IP SNR evolution; the RP process is followed after IP process.  
Figure 2 illustrates the scenario of the spectral evolution of SNRs; how RP is followed after IP.   % 

\noindent
{\bf Phase 1 (IP process)}:~The SNR evolution starts by a shock-heating, 
in which $kT_{\rm i}$ and $kT_{\rm e}$ are nearly 0 keV.
The $kT_{\rm e}$ quickly increases to several keV, which gradually ionize neutral atoms, and hence  $kT_{\rm i}$ also gradually increases following $kT_{\rm e}$ (phase 1).  Thus  phase 1 is ionizing plasma  (IP) process of $kT_{\rm e} > kT_{\rm i}$.
To evolve into the next phase (phase 2) of RP process of $kT_{\rm e} < kT_{\rm i}$, 
there should be a transition phase shown by the epoch A and B in figure 2.   
In this transition phase, $kT_{\rm e}$ drops down below $kT_{\rm i}$, 
by either conductive cooling by cold cloud (conduction: \cite{Kawasaki2002}) 
or adiabatic cooling by break-out of the plasma in a dense medium to a thin medium (rarefaction; \cite{Masai1994}) (figure 2a). 
The other possibility is that $kT_{\rm i}$ increases by either photo-ionization of an external X-ray source, or ionization by low-energy cosmic rays 
(LECRs, e.g., \cite{Nakashima2013}) (figure 2b) \footnote {Depending on the transition process, 
the duration  (A$\rightarrow$B) is not always short as is given in figure 2. In some cases (LECR origin), 
its duration is very long extending from the initial to the present epoch.}.   
 
\noindent
{\bf Phase 2 (RP process)}: After the transition, SNR evolution enters RP process, 
in which free electrons in lower temperature of $kT_{\rm e}$ are recombining to bound states of ions 
in higher ionization temperature of $kT_{\rm i}$. 
This process makes radiative recombining continuum (RRC). 
The RRC structure is direct evidence for RP, and has been clearly observed in Si and S in IC 443 \citep{Yamaguchi2009}. 
As is noted in section 1,
the conventional model of the RP SNRs ignores the phase 1 process and treats only the phase 2 process; 
RP starts from a plasma of $kT_{\rm i} > kT_{\rm e}$, where $kT_{\rm i}$ are all the same in the relevant elements at the epoch B
(here, single-kT$_{\rm i}$). 
Our new model explicitly assumes element dependent  $kT_{\rm i}(z)$ (here, multi-kT$_{\rm i}$(z) model), 
which can be smoothly connected to the IP (phase1), well established  process for almost all of SNRs.

\begin{figure*}[t] %figure 2
\begin{center}
\includegraphics[width=6cm, angle=90]{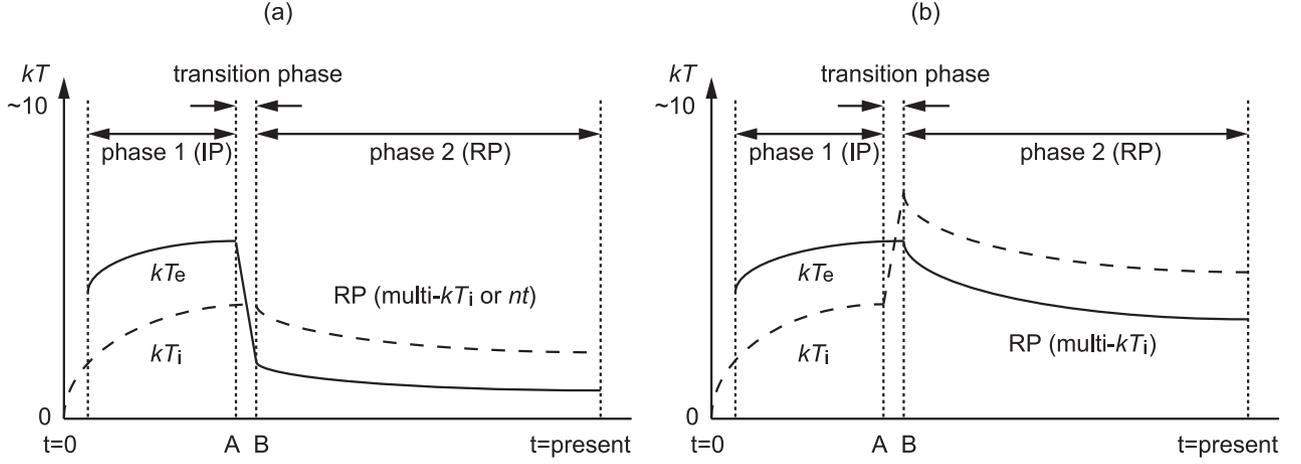}  %IP+RPevolution
\end{center}
\caption{Schematic pictures of the $kT_{\rm e}$ and  $kT_{\rm i}$ evolutions in the RP SNR. 
(a) is the case of electron temperature ($kT_{\rm e}$) decreases, 
and (b) is the case of ionization temperature ($kT_{\rm i}$) increases. 
}
\end{figure*}

The multi-kT$_{\rm i}$(z) model fit has been partially applied to the RP analysis of W28 in the limited energy band (0.6--5 keV) 
with relevant elements of Ne--Fe \citep{Sawada2012}. 
For more comprehensive multi-kT$_{\rm i}$(z) model fit of IC 443 than those of the W28 study, 
we utilized the wider energy band spectrum of 0.6--10~keV. 
This energy band covers essentially all the relevant elements from Ne to Ni.

\subsection{Outline of analysis} %section 3.3

The spectrum of the Background (BGD)  is extracted  
from the nearby sky region; the model spectrum
is based on 
\citet{Masui2009}, which is consisted of the Milky way halo (MWH), 
local hot bubble (LHB) and cosmic X-ray background (CXB) \citep{Kushino2002}. 
For the study of the spectra and fluxes of IC 443, 
we use the data within a circle of a \timeform{6.5'} radius  
centered on (R.A., Dec.)$_{\rm J2000.0}$=(6:17:10.000, $+$22:45:10.00) excluding calibration sources.
IC 443 is located at the anti-center region with the Galactic latitude of $b$=$3^{\circ}$, 
and hence the Galactic ridge X-ray emission (GRXE) is ignored \citep{Uchiyama2013,Yamauchi2016,Koyama2018}. 
Since we find an Fe\emissiontype{I} K$\alpha$ line, which should be associated with a continuum, we add a power-law component (PL) 
with an Fe\emissiontype{I} K$\alpha$ line. 

Thus the formula of the spectral fitting is the sum of plasma from interstellar medium ($ISM$) and SNR ejecta ($Ejecta$), power law ($PL$) plus Fe\emissiontype{I} K$\alpha$ line ($Gaussian\ (6.4 keV)$), and nearby X-ray background ($BGD$). This is given as,
\begin{equation}
  ISM + Ejecta + PL + Gaussian\ (6.4 keV) +BGD.  %(1)
\end{equation}
The ejecta plasma model ($Ejecta$)  is either a single-kT$_{\rm i}$ (conventional model) or 
multi-kT$_{\rm i}$(z) (new model) in the {\tt VVRNEI} code of the XSPEC package (version 12.9.1).
The latter is given as,
\begin{equation}
\sum_{\rm z=H}^{\rm z=Ni}VVRNEI [multi-kT_{\rm i}(z)], %(2)
\end{equation} 
where $kT_{\rm i}(z)$ is variable ionization temperature for each element. 
We added two Gaussians at $\sim$0.8 keV and $\sim$1.2 keV to represent the features due to  
incomplete atomic data for the Fe-L shell complex in the current plasma model (e.g., \cite{Nakashima2013}).

As is given in table 1, the observations are extended over large time span. 
Although the energy resolution in  the first observation is not degraded by particle background, 
those of the latter observations are significantly degraded. 
In order to compensate this energy resolution variations in the summed spectrum, we apply {\tt gsmooth} code in XSPEC. 
The line broadening due to the time-dependent variations of the energy resolution are $\sim$30 eV (FWHM). 
The energy scale is fine-tuned by applying artificial {\tt redshift} for every element. 
The gain variation with energy is a concaved function with the amplitude of $+$1.4\%$\sim$$-$0.4\% in {\tt redshift}.

\subsection{RP model fit} % section 3.4

{\bf Model A}:~ At first, we fit the IC 443 spectrum using equation (1) of a single-kT$_{\rm i}$ model (RP1) for the
ejecta spectrum ($Ejecta$). Free parameters are $N_{\rm H}$, $kT_{\rm e}$, $kT_{\rm i}$ ($kT_{\rm i}(z)$), recombination timescale ($n_{\rm e}t$), 
normalizations and abundances of Ne--Ni, while abundances of He, C, N, and O are assumed to be 1 solar. 
The abundance tables and the atomic data of the lines and continua of the thin thermal plasma are taken from 
\citet{Anders1989} and ATOMDB 3.0.7, respectively. 

The best fit $kT_{\rm e}$ of $ISM$ is $\sim0.2$ keV, for which 
there is no significant difference between CIE and non-equilibrium ionization (NEI). 
We, therefore, assume that $ISM$ is CIE with the solar abundance (see e.g., \cite{Matsumura2017}). 
The global fit is rejected with a large $\chi^2$/d.o.f. of 2473/938 (2.64).

\noindent
{\bf Model B}:~ Next, we fit  with a multi-kT$_{\rm i}$(z) model (RP1) of equation (2). Free parameters  
are same as model A. The global fit is significantly improved with $\chi^2$/d.o.f. of 2007/932 (2.15). 
However, we find significant residuals in the 5--10 keV band, in particular at the Fe \emissiontype{I} K$\alpha$ (6.4 keV),
Fe\emissiontype{XXV} He$\alpha$ (6.7 keV), Ni \emissiontype{VVVII} He$\alpha$ (7.8 keV), and at RRC of Fe\emissiontype{XXV} (8.83 keV).

\noindent
{\bf Model C}:~We, therefore, add another {\tt VVRNEI} model (RP2) and a PL component. 
The {\tt VVRNEI} model (RP2)  is closely related to the He$\alpha$ and Ly$\alpha$ lines of Cr--Ni.
The normalization and the recombination timescale for RP2 are linked to those of the lower-temperature component (RP1).
The PL is associated with an Fe\emissiontype{I} K$\alpha$ line.  
We cannot constrain the photon index ($\Gamma$) value by the fitting. 
Non-thermal SNRs typically have $\Gamma$ of 2--3 (e.g., SN1006: Bamba et al. 2003, 2008; RX J1713-3946: Koyama et al. 1997). 
Accordingly, we assume the power-law index $\Gamma$ to be 2.5. 
The line width is assumed to be 0 keV. 
The global fit is largely improved with $\chi^2$/d.o.f. of 1674/923 (1.81).  
In the pure statistical point of view, model C is still unacceptable. 
However, possible errors, such as atomic data of  L-lines of Fe\emissiontype{XVI}   and Fe\emissiontype{XVIII}, 
and small calibration errors near the Si\emissiontype{I} K-edge energy,  may not be negligible for  
the spectrum with very high photon statistics. 
The actual photon count rate in the Si K-edge band (1.83--1.85 keV) is $(4.20\pm0.02)\times 10^{-2}$ counts s$^{-1}$. 
Taking account of these systematic errors, we use the model C results as a reasonable approximation. 
The best-fit spectra of model A, B, and C are given in figure 3, while the best-fit parameters of model A, B, and C are listed in table 2. 
The $kT_{\rm i}(z)$ monotonously increases as $z$ in Ne--Ca, then decreases in Cr--Ni for the RP1 component. 
The abundances of Ne--Ni are generally moderate. 
Some fractions of the Cr--Ni plasma (RP2) show a drastic increase of $kT_{\rm i}(z)$. 
In the Cr--Ni plasma, the abundance ratio of Ni/Fe is $\sim$5, 
but the abundances of Cr--Ni are very small.  

\begin{figure*} [tbh] % Figure 3
\begin{center}
\includegraphics[width=8cm, angle=0]{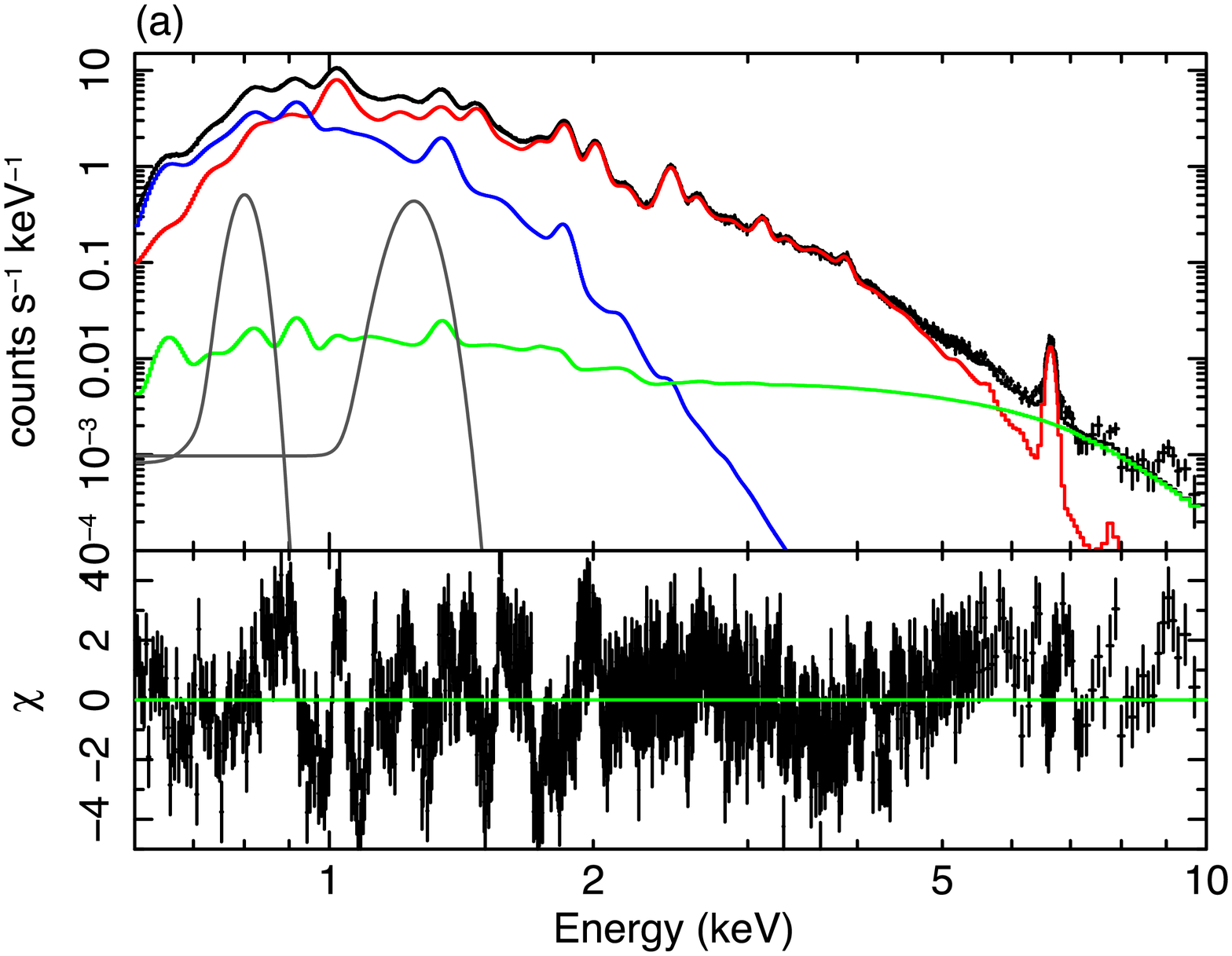} %model A single kTi
\includegraphics[width=8cm, angle=0]{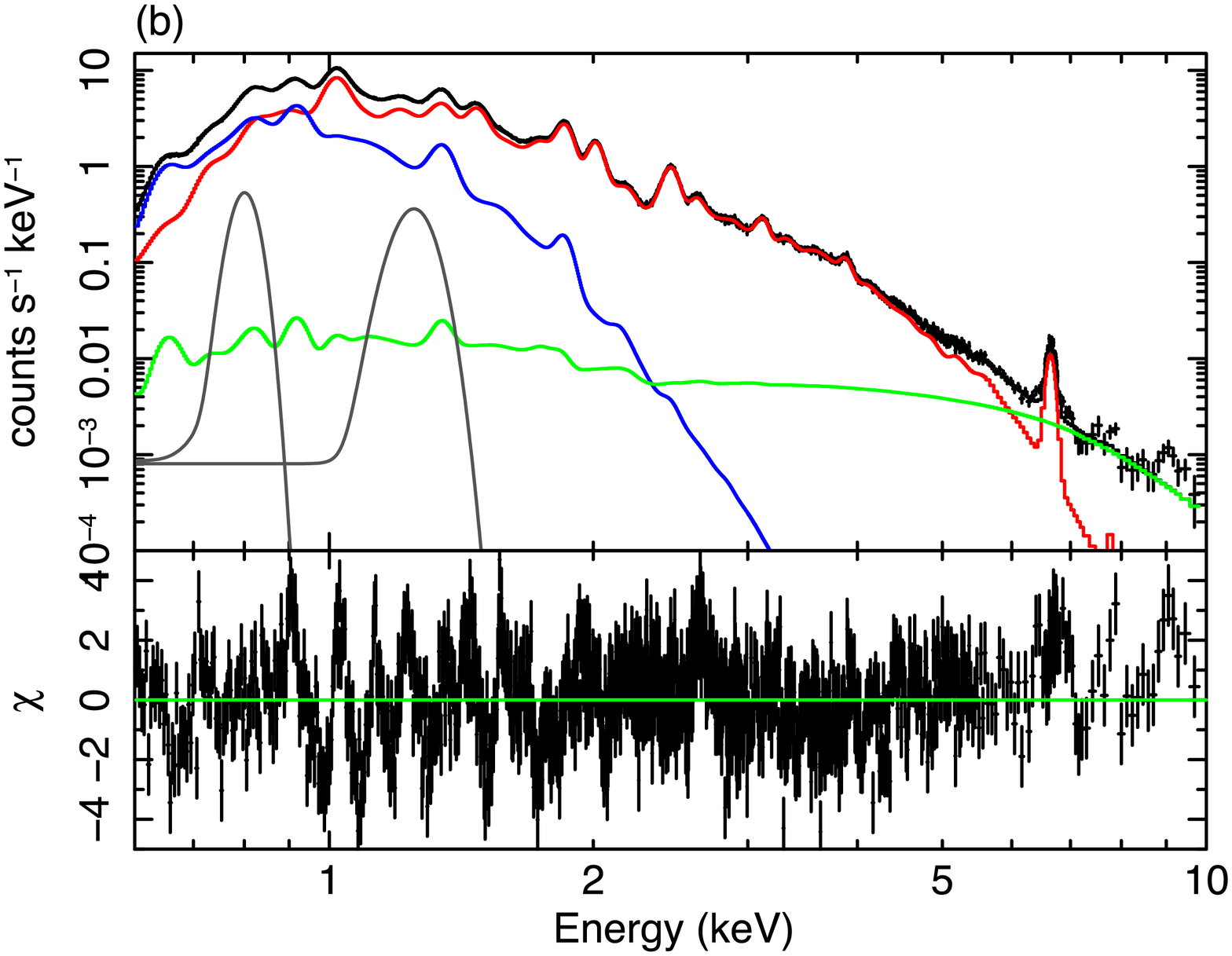} %model B 1-RP multi-kTi
\includegraphics[width=8cm, angle=0]{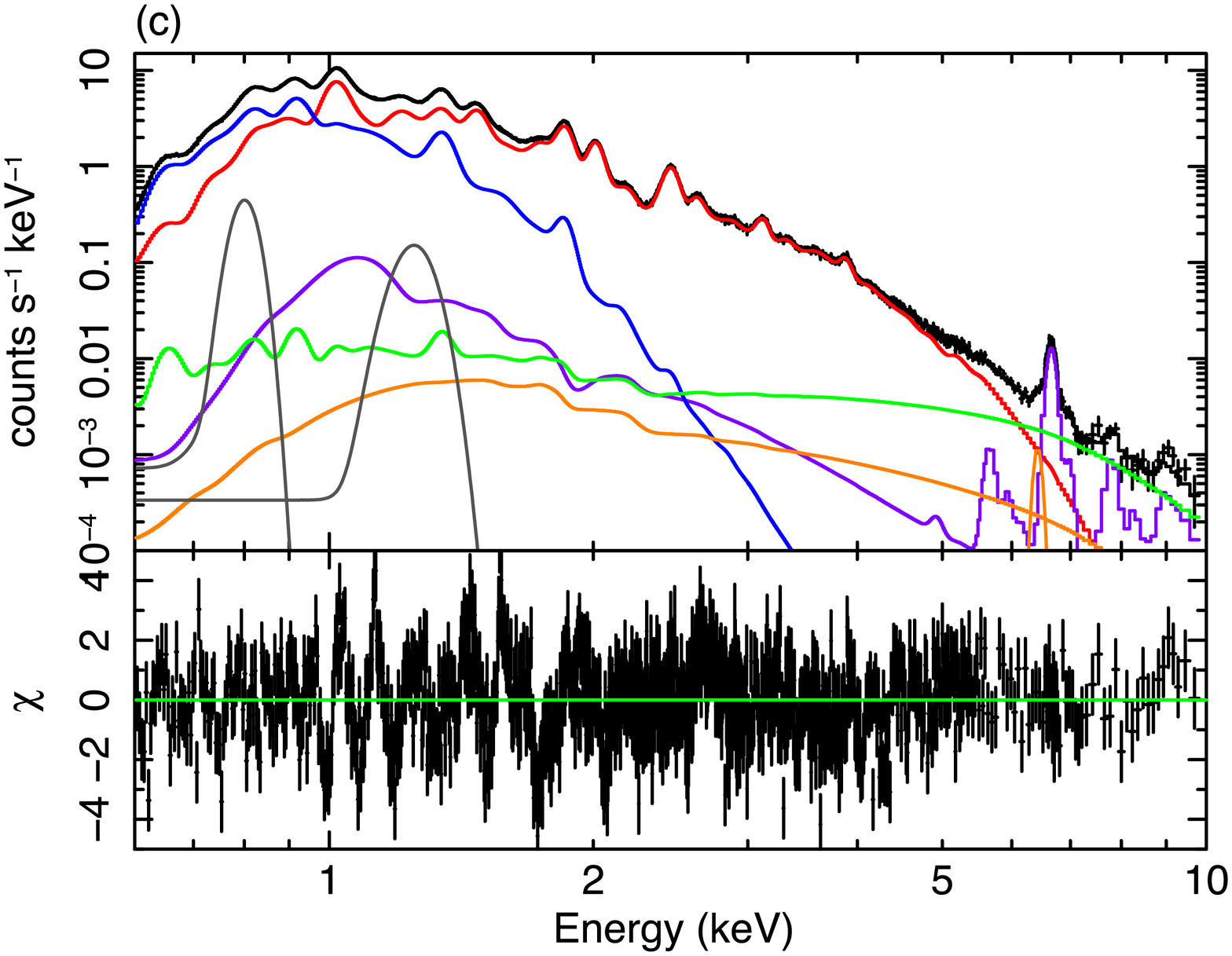} %model C 2-RP multi-kTi
\includegraphics[width=8cm, angle=0]{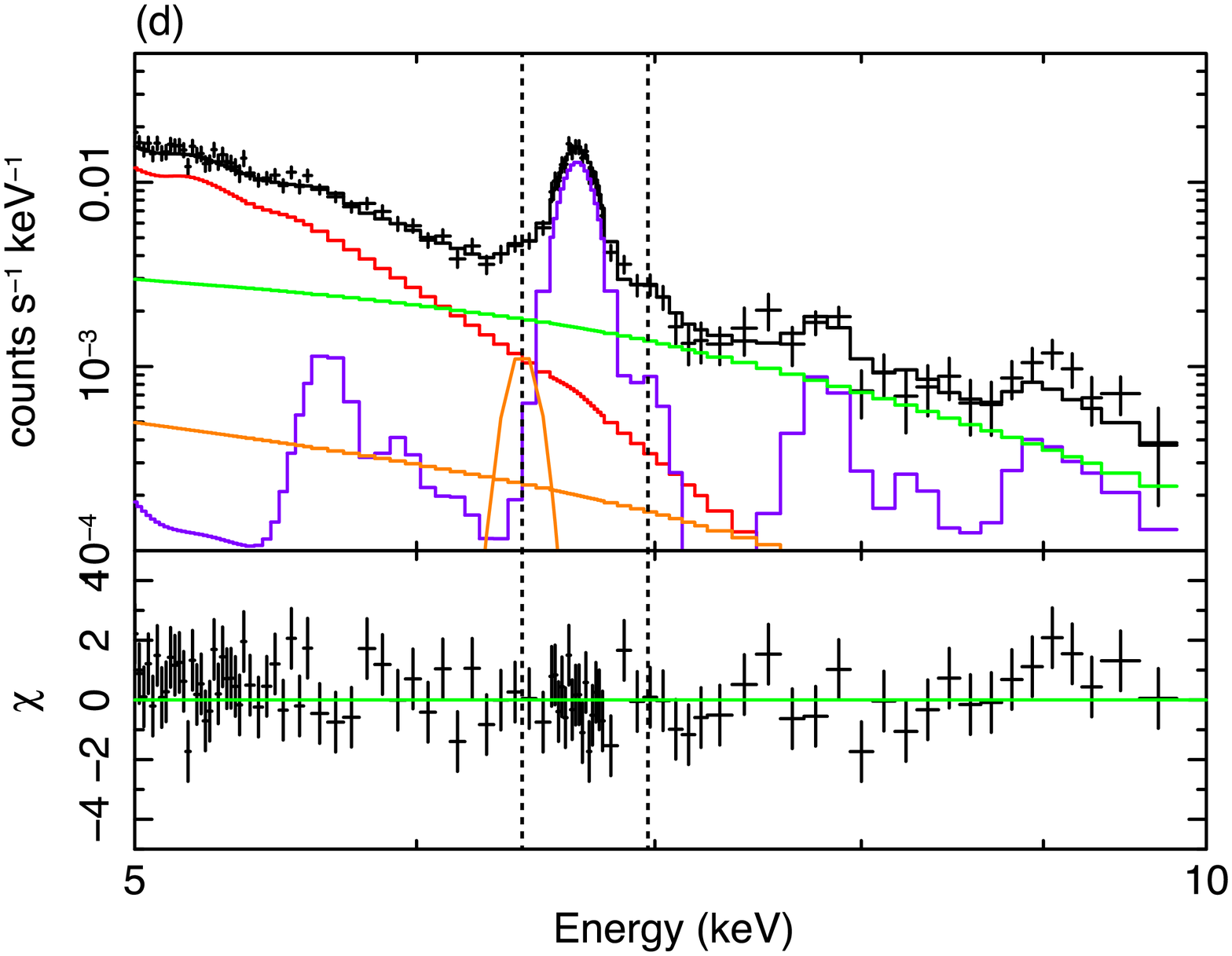} %model C 2-RP multi-kTi
\end{center}
\caption{The XIS spectrum of IC 443 and the best-fit models: (a) single-kT$_{\rm i}$ (model A), (b) multi-kT$_{\rm i}$(z) (model B) 
(c) two-RP of multi-kT$_{\rm i}$(z) (model C), and (d) same as (c) but in the 5--10 keV band.
The solid red, purple, blue, and orange lines are RP1, RP2, CIE, and PL plus an Fe \emissiontype{I} K$\alpha$ line, respectively.
The gray lines show Fe-L lines, 
while the green line shows the BGD (=MWH$+$LHB$+$CXB). 
Dotted lines in (d) indicate data excess at the energies of Fe \emissiontype{I} K$\alpha$ (left) and Fe \emissiontype{XXVI} Ly$\alpha$ lines (right). }
\end{figure*}

\begin{table*}
\footnotesize
\caption{The best-fit parameters of  single-kT$_{\rm i}$ (model A),  multi-kT$_{\rm i}$(z) (model B), two-RP of  multi-kT$_{\rm i}$(z) (model C), and 
 two-RP of  multi-kT$_{\rm i}$(z) in the CC-SNR case (model D).}
 \begin{center}
 \begin{tabular}{llcccccccc} \hline
& & \multicolumn{2}{c}{model A} & \multicolumn{2}{c}{model B} & \multicolumn{2}{c}{model C} & \multicolumn{2}{c}{model D} \\ %\hline
Component & Parameter & \multicolumn{8}{c}{Values} \\ \hline
Absorption& %\multicolumn{6}{c}{Absorption} \\ \hline
$N_{\rm H}^{\ast}$ &%($\times$10$^{22}$ cm$^{-2}$)& 
\multicolumn{2}{c}{0.65$\pm$0.01} &  \multicolumn{2}{c}{0.65$\pm$0.01}  &  \multicolumn{2}{c}{0.67$\pm$0.01}  &  \multicolumn{2}{c}{0.67$\pm$0.01}  \\ \hline
$ISM$\ (CIE) & $kT_{\rm e}^{\dag}$ & 
\multicolumn{2}{c}{0.22$\pm$0.01} &  \multicolumn{2}{c}{0.21$\pm$0.01}  &  \multicolumn{2}{c}{0.22$\pm$0.01}  &  \multicolumn{2}{c}{(= model C)}  \\
& $VEM^{\ddag}$ & 
\multicolumn{2}{c}{1.7$\pm$0.7} &  \multicolumn{2}{c}{1.8$\pm$0.1}  &  \multicolumn{2}{c}{2.0$\pm$0.1}  &  \multicolumn{2}{c}{(= model C)}  \\ \hline
$Ejecta$\ (RP1) & $kT_{\rm e}^{\dag}$ & 
\multicolumn{2}{c}{0.53$\pm$0.01} &  \multicolumn{2}{c}{0.53$\pm$0.01}  &  \multicolumn{2}{c}{0.56$\pm$0.01}  &  \multicolumn{2}{c}{(= model C)}  \\
%$kT_{\rm i}$1 (keV) & \multicolumn{2}{c}{0.222$\pm$0.001} &  \multicolumn{2}{c}{---}  &  \multicolumn{2}{c}{---}  \\
& $VEM^{\ddag}$ & \multicolumn{2}{c}{0.47$\pm$0.02} &  \multicolumn{2}{c}{0.54$\pm$0.03}  &  \multicolumn{2}{c}{0.46$\pm$0.04}  &  \multicolumn{2}{c}{0.042$\pm$0.001} \\      
& $n_{\rm e}t^{\S}$ %($\times$10$^{11}$ cm$^{-3}$ s) 
   & \multicolumn{2}{c}{3.6$\pm$0.1} &  \multicolumn{2}{c}{2.9$\pm$0.2}  &  \multicolumn{2}{c}{2.0$\pm$0.1}  &  \multicolumn{2}{c}{(= model C)}  \\ 
&   & $kT_{\rm i}(z)^{\dag}$ & $Ab^{\Vert}$  & $kT_{\rm i}(z)^{\dag}$ & $Ab^{\Vert}$ & $kT_{\rm i}(z)^{\dag}$ & $Ab^{\Vert}$ & $kT_{\rm i}(z)^{\dag}$ & $Ab^{\Vert}$ \\  
&    Ne & 2.0$\pm$0.1  & 2.9$\pm$0.3 	& 0.94$\pm$0.12 & 2.6$\pm$0.2 		& 0.85$\pm$0.05& 3.4$\pm$0.5        & 0.73$\pm$0.02 & 34$\pm$2 \\
&    Mg & (link to Ne)  	& 1.8$\pm$0.2 		& 1.1$\pm$0.1 & 1.5$\pm$0.1  		& 1.1$\pm$0.1 & 1.6$\pm$0.1		& (= model C) & 19$\pm$1 \\
&    Al & (link to Ne) 	& 2.2$\pm$0.3 		& (link to Mg) & 1.3$\pm$0.2		& (link to Mg) & 1.1$\pm$0.2		& (= model C) & 15$\pm$2 \\
&    Si & (link to Ne)  	& 2.1$\pm$0.2 		& 1.7$\pm$0.1 & 1.9$\pm$0.1 		& 1.5$\pm$0.1 & 2.0$\pm$0.2		& (= model C) & 22$\pm$1 \\
&    S & (link to Ne)  	& 2.5$\pm$0.2 		& 1.7$\pm$0.1 & 2.2$\pm$0.1 		& 1.6$\pm$0.1 & 2.3$\pm$0.2		& (= model C) & 26$\pm$1 \\
&    Ar & (link to Ne)  	& 2.2$\pm$0.3 		& 1.9$\pm$0.1 & 1.8$\pm$0.2		& 1.6$\pm$0.1 & 1.9$\pm$0.2		& (= model C) & 21$\pm$1 \\
&    Ca & (link to Ne)  	& 3.2$\pm$0.4 		& 3.9$\pm$0.4 & 0.71$\pm$0.10 	& 2.7$\pm$0.4 & 0.92$\pm$0.22		& (= model C) & 10$\pm$1 \\
&    Cr=Mn & (link to Ne)  & 10$\pm$3 	& (link to Fe) & 12$\pm$2 			&  (link to Fe) & $<$0.4			&  (= model C) & $<$8 \\
&    Fe & (link to Ne)  	& 0.47$\pm$0.29 	& 1.4$\pm$0.1 & 0.45$\pm$0.02	&0.81$\pm$0.09 & 0.33$\pm$0.03	& (= model C) & 3.8$\pm$0.2 \\
&    Ni & (link to Ne)  	& 2.3$\pm$0.2 		& (link to Fe) & 1.7$\pm$0.2 		& (link to Fe) & 2.3$\pm$0.2		& (= model C) & 28$\pm$2 \\
\hline
$Ejecta$\ (RP2) & $kT_{\rm e}^{\dag}$ & \multicolumn{2}{c}{---} &  \multicolumn{2}{c}{---}  &  \multicolumn{2}{c}{0.64$\pm$0.11}   &  \multicolumn{2}{c}{(= model C)} \\
& $VEM^{\ddag}$ & \multicolumn{2}{c}{---} &  \multicolumn{2}{c}{---}  &  \multicolumn{2}{c}{0.46 (link to RP1)}  &  \multicolumn{2}{c}{0.042 (link to RP1)}\\      
& $n_{\rm e}t^{\S}$ %($\times$10$^{11}$ cm$^{-3}$ s)  
& \multicolumn{2}{c}{---} &  \multicolumn{2}{c}{---}  &  \multicolumn{2}{c}{2.0 (link to RP1)}     & \multicolumn{2}{c}{(= model C)} \\ 
&    &  &   &  &  & $kT_{\rm i}^{\dag}$ & $Ab^{\Vert}$ & $kT_{\rm i}^{\dag}$ & $Ab^{\Vert}$ \\  
&    Cr & --- & --- & --- & --- & (link to Fe) & 0.30$\pm$0.17  & (link to Fe) & 3.4$\pm$1.6 \\
&    Mn & --- & --- & --- & --- & (link to Fe) & $<$0.5  & (link to Fe) & $<$5.0 \\
&    Fe  & --- & --- & --- & --- & 6.2$\pm$0.4 & 0.030$\pm$0.003  & (= model C) & 0.33$\pm$0.02 \\
&    Ni  & --- & --- & --- & --- & (link to Fe) & 0.16$\pm$0.10  & (link to Fe) & 1.7$\pm$1.0  \\
\hline
$PL$ & $\Gamma^{\sharp}$ & \multicolumn{2}{c}{---} &  \multicolumn{2}{c}{---}  &  \multicolumn{2}{c}{2.5 (fixed)}  &  \multicolumn{2}{c}{2.5 (fixed)}  \\
& $Norm^{\sharp}$ & 
\multicolumn{2}{c}{---} &  \multicolumn{2}{c}{---}  &  \multicolumn{2}{c}{7.8$^{+10.1}_{-2.8}$}  &  \multicolumn{2}{c}{(= model C)}\\ \hline
$Gaussian$  
& $E^{\ast \ast}$ &%energy &%(keV) & 
\multicolumn{2}{c}{---} & \multicolumn{2}{c}{---} & \multicolumn{2}{c}{6.43$\pm$0.04}  & \multicolumn{2}{c}{(= model C)}  \\
& $F_{\rm 6.4 keV}^{\ast \ast}$ & 
\multicolumn{2}{c}{---} & \multicolumn{2}{c}{---} & \multicolumn{2}{c}{6.7$\pm$3.0}   & \multicolumn{2}{c}{(= model C)} \\\hline
%\\ \hline
& $\chi^2$/d.o.f. & \multicolumn{2}{c}{2473/938=2.64} &  \multicolumn{2}{c}{2007/932=2.15}  &  \multicolumn{2}{c}{1674/923=1.81}  &  \multicolumn{2}{c}{1722/941=1.83}  \\     
     \hline\\
   \end{tabular}
   \end{center}
$^{\ast}$ The unit is $\times$10$^{22}$ cm$^{-2}$. \\
$^{\dag}$ Units of $kT_{\rm e}$ and $kT_{\rm i}(z)$ are keV. \\
$^{\ddag}$ Volume emission measure ($VEM$=$n_{\rm e}n_{\rm H}V$) with the distance of 1.5 kpc. The unit is $10^{58}$cm$^{-3}$. \\
$^{\S}$ Recombination time scale, where $n_{\rm e}$ is the electron density (cm$^{-3}$) and $t$ is the elapsed time (s). The unit is $\times$10$^{11}$ cm$^{-3}$ s. \\
$^{\Vert}$ Relative to the solar values in \citet{Anders1989}. \\
$^{\sharp}$ Power-law index and flux normalization. The unit of normalization is 10$^{-7}$ photons s$^{-1}$ cm$^{-2}$ keV$^{-1}$  arcmin$^{-2}$ at 1 keV. \\
$^{\ast \ast}$ The Fe \emissiontype{I} K$\alpha$ line energy and photon flux in units of keV and 10$^{-9}$ photons s$^{-1}$ cm$^{-2}$ arcmin$^{-2}$, respectively.
\normalsize
\end{table*}

\section{Discussion} % section 4

\subsection{Two-RP model and its implication} % section 4.1

The IC 443 spectrum is nicely fitted by a two-RP (RP1 and RP2) of multi-kT$_{\rm i}$(z) model. 
The best-fit $kT_{\rm e}$ of RP1 and RP2 are 0.56 and 0.64 keV, respectively.
\citet{Matsumura2017} also obtained nice fit with a 
similar two-RP model for IC 443. 
However, the best-fit $kT_{\rm e}$ are 0.24 and 0.61 keV, respectively, largely different from our best-fit model. 
The best-fit abundances are also about two times larger than ours. 
We note that \citet{Matsumura2017} used  limited data set from ID=501006010 (table 1) and limited energy band of 0.6--7.5 keV. 
We, therefore, re-fit their model  to our  higher quality data (includes all the Obs. ID in table 1) in the wider energy band (0.6--10 keV) to include the RRC of Fe. Then, the fit is rejected with $\chi^2$/d.o.f.=3589/926=3.88, which is larger than our simpler model A  
of 1-$kT_{\rm e}$ and 1-$kT_{\rm i}$  of $\chi^2$/d.o.f.= 2473/938=2.64. 
We also compare our best-fit results with the other previous IC 443 report, and find that model C is the best in reliability and quantity. 

In the multi-kT$_{\rm i}$(z) model (model C) of ejecta, we found the derived $kT_{\rm i}(z)$ for each element is not the same, but shows monotonous increase in Ne--Ca and decrease in Cr--Ni for RP1 (table 2).
Since RP2 has different electron and ionization temperatures from those of RP1, the RP2 plasma may have different origin. 
The RP2 component shows a higher Ni abundance than Fe (table 2), and might be dominated by the spectrally-hard emission concentrated in the central region of 
the SNR. Although it is speculative, this hard component emission appears to involve X-ray emission from elongated regions connected to the brightest SNR 
center (green contours in figure 1) in contrast to the largely smooth distribution of the spectrally-softer RP1 component (dotted contours in figure 1).
We speculate that the RP2 component emission might be associated with strong outflows of neutron-rich gas in the deepest core of the CC-SN. 
Deep follow-up high-resolution X-ray imaging spectroscopic observations of this central region of IC 443 would be required to test this scenario.

In model A-C (table 2), we fit the ejecta plasma with fixed abundances of He--O of 1 solar, because these elements do not explicitly appear in the spectrum of $\gtsim$0.7 keV. Since CC-SNRs are largely over-solar abundances of He--O, we re-fit abundances for the case of He$\sim$2.6, C$\sim$6.4, N$\sim$6.5, and O$\sim$28 solar (model D), the mean values of \citet{Woosley1995}. 
The results are given in table 2. The re-fitted abundances in RP1 are about 10 times larger than those of model C. 
The very small abundances of Fe and Ni in RP2 of model C increase to $\sim$0.3 and $\sim$1.7 solar in model D, respectively.  
The abundances for individual elements and their distribution in the RP1 and RP2 plasma are roughly consistent with the initial assumption of CC-SNR (e.g., \cite{Woosley1995}).

From the $kT_{\rm i}$ given in table 2, the ionization temperature of Si and S in IC 443 is estimated to be $\sim$1.5 keV at the start of RP process (epoch A in figure 2). This temperature gives significant Si\emissiontype{XIV} Ly$\alpha$ and S\emissiontype{XVI} Ly$\alpha$ as is given in figure 3c. 
We check the available SNR spectra with similar $kT_{\rm e}$ of IC 443 ($kT_{\rm e}\sim$0.3--1 keV), but find no significant Si\emissiontype{XIV} Ly$\alpha$ and S\emissiontype{XVI} Ly$\alpha$ from the IP SNR samples (e.g., Cygnus loop: \cite{Uchida2011}; G272.2$-$3.2: \cite{Kamitsukasa2016}; Kes 79: \cite{Sato2016}). 
As for Fe\emissiontype{XXVI} Ly$\alpha$, the IP SNR samples are limited, but 
the good sample is Cas A, which shows no Fe\emissiontype{XXVI} Ly$\alpha$ ($kT_{\rm e}$=1--2 keV, \cite{Hwang2012}). 
On the other hand, all the spectra of RP SNR samples, which corresponds to the phase 2 (after the transition of IP$\rightarrow$RP), have clear Si\emissiontype{XIV} Ly$\alpha$ and S\emissiontype{XVI} Ly$\alpha$ (e.g., W28: \cite{Sawada2012}; G359.1$-$0.1: \cite{Ohnishi2011}). 
Some RP SNRs with better statistics such as IC 443 (this paper) and W49B \citep{Ozawa2009} show clear Fe\emissiontype{XXVI} Ly$\alpha$ line. 
Since all the RP SNRs would be made by the transition of IP$\rightarrow$RP (phase 1$\rightarrow$phase 2), the observed Si\emissiontype{XIV} Ly$\alpha$, S\emissiontype{XVI} Ly$\alpha$, and Fe\emissiontype{XXVI} Ly$\alpha$ in IC 443 should be made by this transition. 

The conventional model of electron cooling (figure 2a) does not increase the ionization temperature to make significant Si\emissiontype{XIV} Ly$\alpha$, S\emissiontype{XVI} Ly$\alpha$, and Fe\emissiontype{XXVI} Ly$\alpha$ lines, and hence cannot change the IP to RP in the transition phase. The electron cooling by thermal conduction to cool clouds has another problem that an ionization temperature would  also decrease by a contamination of cloud evaporation, 
and hence the plasma becomes even IP, not RP in the transition phase of figure 2a.
Therefore, we rather prefer the model of ionization temperature increase shown in figure 2b, 
in which ionization temperature at the epoch B can be high enough to make Ly$\alpha$ lines of Si, S, and Fe and Fe-RRC. The underlying process would be an irradiation of external X-rays or LECRs.

We discover an Fe\emissiontype{I} K$\alpha$ line (6.4 keV) associated with a PL component. In order to explain both the RP and Fe\emissiontype{I} K$\alpha$ line, 
we propose a new model which is out of the standard scheme of SNR evolution. LECRs are successively produced by the SNR shock. They irradiate the SNR hot plasma and nearby cold cloud. Fe\emissiontype{XXV} in the hot plasma is ionized to Fe\emissiontype{XXVI}, then free electrons with a temperature of $\sim$0.6--0.7 keV in the hot plasma are recombined to the ground state of Fe\emissiontype{XXVI}  
after a long recombination process of $(2.0\pm0.1)\times10^{11}$ cm$^{-3}$ s. This process makes the Fe-RRC structure. 
For the origin of Fe\emissiontype{I} K$\alpha$ line, one may argue that the origin is irradiation on molecular cloud by an external X-ray source. 
However this is remote possibility, because no bright X-ray source is found near IC 443. We propose another possibility that the origin of Fe\emissiontype{I} K$\alpha$ plus continuum is due to LECRs. Our speculation, therefore, 
is that LECRs are successively produced by the SNR shock, and irradiate both the SNR hot plasma (responsible for RP) and nearby cold cloud (responsible for PL plus Fe\emissiontype{I} K$\alpha$). 

\subsection{A possible problem of the RP code} % section 4.2

The model C including {\tt VVRNEI} in XSPEC, can not fully explain the fluxes of Fe \emissiontype{XXV} He$\alpha$, Fe\emissiontype{XXVI} Ly$\alpha$ and
 Fe\emissiontype{XXV}-RRC in the RP2 spectrum 
(see 5--10 keV band in figure 3d) in IC 443; 
the observed flux ratio of RRC/He$\alpha$ is larger than code prediction. 
On the other hand, the RP code of SPEX for the model C successfully reproduces the RRC/He$\alpha$ ratio
in the 3.7--10 keV band spectrum \citep{Ohnishi2014}. 
We therefore try to re-fit the 5--10 keV band spectrum using the SPEX code of the latest version. 
Unlike the XSPEC code, this model using the SPEX well reproduces the 5--10 keV band spectrum, 
in particular the flux ratio of Fe-RRC/Fe\emissiontype{XXV} He$\alpha$. 
Thus, we confirm a problem of disagreement between XSPEC and SPEX. 
To investigate the origin of this code disagreement  in the Suzaku spectrum is beyond a scope of this paper. 
The high resolution calorimeter on board future mission of XRISM may solve this question, 
because it can resolve the fine structure of Fe He$\alpha$ ($x, y, z$, and $w$) and Ly$\alpha$1 and Ly$\alpha$2. 
The absolute flux and the flux ratios of these fine structure lines are essential for the basic atomic process in the RP 
and the the other plasma processes in SNR.

\section{Conclusion}
We provide the IC 443 spectrum of the highest signal-to-noise ratio in the widest energy band of 0.6--10 keV.
The analysis process and results for the unprecedented spectrum are given as follows.      

\begin{itemize}

\item  The multi-kT$_{\rm i}$(z) model of RP is nicely fitted to the IC 443 spectrum.  

\item The distribution of $kT_{\rm i}(z)$ as $z$ in Ne--Ca monotonously increases, then decreases in Cr$-$Ni. However, a fraction of Cr--Ni show a drastic increase.

\item  Cr\emissiontype{XXIII} He$\alpha$ and Fe\emissiontype{I} K$\alpha$ lines are found for the first time.

\item  The high abundance ratio of Ni/Fe may support that Ni is over-produced in the neutron rich central region (in the neutron star).

\item  We interpret that the origin of the RP and an Fe\emissiontype{I}  K$\alpha$ line in IC 443 are enhanced ionization by an irradiation of LECRs.
 
\end{itemize}

\section*{Acknowledgement}

The authors are grateful to all members of the Suzaku team. 
This work was supported by the Japan Society for the Promotion of Science (JSPS) 
KAKENHI Grant Numbers JP16J00548 (KKN), JP15H02090, JP17K14289 (MN) and Nara Women's University Intramural Grant for Project Research (SY).

\end{document}